\documentclass[prl,aps,preprint,showpacs,superscriptaddress,floatfix]{revtex4-1}

\usepackage{epsf}
\usepackage{bm}
\usepackage{amsmath}
\usepackage{amssymb}
\usepackage{amsgen}
\usepackage{amsfonts}
\usepackage{amsbsy}
\usepackage{version}
\usepackage{url}
\usepackage{wasysym}
\usepackage[dvips]{epsfig,color}
\usepackage{psfrag}
\usepackage{mathdots}

\begin{document}
\newcommand{\rmd}{\mathrm{d}}
\newcommand{\Tr}{\mathrm{Tr}}

\renewcommand{\vec}[1]{\bm{#1}}
\renewcommand{\hat}[1]{\bm{\widehat{#1}}}

\renewcommand{\Re}{\mathrm{Re}\,}
\renewcommand{\Im}{\mathrm{Im}\,}

\newcommand{\red}[1]{{\color{red}#1}}
\newcommand{\green}[1]{{\color{green}#1}}
\newcommand{\blue}[1]{{\color{blue}#1}}
\newcommand{\cyan}[1]{{\color{cyan}#1}}
\newcommand{\magenta}[1]{{\color{magenta}#1}}

\newcommand{\fulltriangleup}{\mbox{$\blacktriangle$}}
\newcommand{\fulltriangledown}{\mbox{$\blacktriangledown$}}
\newcommand{\fullcircle}{\mbox{{\Large$\bullet$}}}
\newcommand{\fullsquare}{\mbox{$\blacksquare$}}
\newcommand{\opentriangle}{\mbox{$\triangle$}}
\newcommand{\opencircle}{\mbox{\Large$\circ$}}
\newcommand{\opensquare}{\mbox{$\Box$}}

\newcommand{\shortdashedline}{\protect\rule[2pt]{1pt}{1pt}\,\,\protect\rule[2pt]{1pt}{1pt}\,\,\protect\rule[2pt]{1pt}{1pt}}
\newcommand{\mediumdashedline}{\protect\rule[2pt]{2pt}{1pt}\,\,\protect\rule[2pt]{2pt}{1pt}\,\,\protect\rule[2pt]{2pt}{1pt}}
\newcommand{\dashedline}{\protect\rule[2pt]{3pt}{1pt} \!\protect\rule[2pt]{3pt}{1pt} \!\protect\rule[2pt]{3pt}{1pt}}
\newcommand{\fullline}{\protect\rule[2pt]{15pt}{1pt}}

\title{The ``magic'' angle in the self-assembly of colloids suspended in a nematic host phase}

\author{Sergej Schlotthauer} 
\affiliation{Stranski-Laboratorium f\"ur Physikalische und Theoretische Chemie,
Fakult\"at f\"ur Mathematik und Naturwissenschaften,
Technische Universit\"at Berlin,
Stra{\ss}e~des~17.~Juni 115, 
10623 Berlin, GERMANY}
\author{Tillmann Stieger} 
\affiliation{Stranski-Laboratorium f\"ur Physikalische und Theoretische Chemie,
Fakult\"at f\"ur Mathematik und Naturwissenschaften,
Technische Universit\"at Berlin,
Stra{\ss}e~des~17.~Juni 115, 
10623 Berlin, GERMANY}
\author{Michael Melle} 
\affiliation{Stranski-Laboratorium f\"ur Physikalische und Theoretische Chemie,
Fakult\"at f\"ur Mathematik und Naturwissenschaften,
Technische Universit\"at Berlin,
Stra{\ss}e~des~17.~Juni 115, 
10623 Berlin, GERMANY}
\author{Marco G. Mazza} 
\affiliation{Max-Planck-Institut f\"ur Dynamik und Selbstorganisation,
Am Fa{\ss}berg 17, 37077 G\"ottingen, GERMANY}
\author{Martin Schoen}
\affiliation{Stranski-Laboratorium f\"ur Physikalische und Theoretische Chemie,
Fakult\"at f\"ur Mathematik und Naturwissenschaften,
Technische Universit\"at Berlin,
Stra{\ss}e~des~17.~Juni 115, 
10623 Berlin, GERMANY}
\affiliation{Department of Chemical and Biomolecular Engineering,
911 Partners Way,
North Carolina State University,
Raleigh, NC 27695, U.S.A.}
\date{\today}

\begin{abstract}
Using extensive Monte Carlo (MC) simulations of colloids immersed in a nematic liquid crystal we compute an effective interaction potential via the local nematic director field and its associated order parameter. The effective potential consists of a local Landau-de Gennes (LdG) and a Frank elastic contribution. Molecular expressions for the LdG expansion coefficients are obtained via classical density functional theory (DFT). The DFT result for the LdG parameter $A$ is improved by locating the phase transition through finite-size scaling theory. We consider effective interactions between a pair of homogeneous colloids with Boojum defect topology. In particular, colloids attract each other if the angle between their center-of-mass distance vector and the far-field nematic director is about $30^{\circ}$ which settles a long-standing discrepancy between theory and experiment. Using the effective potential in two-dimensional MC simulations we show that self-assembled structures formed by the colloids are in excellent agreement with experimental data.
\end{abstract}
\pacs{61.30.-v,61.30.Jf,82.70Dd,05.10.Ln}
\maketitle

If a liquid crystal is in the nematic phase the overall orientation of its molecules (i.e., mesogens) can be described quantitatively by the non-local unit vector (i.e., the far-field nematic director) $\hat{n}_{0}$ \cite{degennes95}. Immersing a colloidal particle in this nematic host phase gives rise to a director field $\hat{n}\left(\vec{r}\right)$ such that sufficiently close to the colloid's surface, $\hat{n}\left(\vec{r}\right)$ and $\hat{n}_{0}$ may differ. The deviation between $\hat{n}\left(\vec{r}\right)$ and $\hat{n}_{0}$ is caused by the specific anchoring of mesogens at the surface of the colloid. Depending on details of the host phase $\hat{n}\left(\vec{r}\right)$ can be of such dazzling complexity that experts are just beginning to unravel its structural details \cite{jampani11}. 

The mismatch between $\hat{n}\left(\vec{r}\right)$ and $\hat{n}_{0}$ also gives rise to effective interactions between several colloids that are mediated by the nematic host \cite{izaki13}. These interactions may therefore be used to self-assemble the colloids into supramolecular entities in a controlled (i.e., directed) manner. This way ordered assemblies of colloids of an enormously complex structure with rich symmetries may be built that would not exist without the ordered structure of the host phase \cite{ognysta09,qi06}. 

The complex self-assembled structures formed by the colloids are also of practical importance. For instance, taking as a specific example dielectric colloids it could be demonstrated  that the propagation of light through a self-assembled ordered colloidal arrangement is affected in a way similar to the propagation of electrons in a semiconductor crystal \cite{humar09}. Hence, ordered periodic assemblies of colloids are already discussed within the framework of novel photonic devices with fascinating properties \cite{musevic11}.

Clearly, to use the effective interaction potential for the self-assembly of colloids in a nematic host phase the molecular origin of the potential itself must be understood. Our motivation to contribute to such an improved understanding  goes back to an observation made some time ago by Poulin and Weitz \cite{poulin98}. They found experimentally that in a nematic phase the colloidal center-to-center distance vector $\vec{r}_{12}$ forms a ``magic'' angle of $\theta\approx30^{\circ}$ with $\hat{n}_{0}$ if the mesogens at the surfaces of the colloidal pair are anchored in a locally planar fashion. Hence, near an isolated colloid a so-called Boojum defect would arise under these conditions \cite{poulin97}.

This experimental observation has resisted a quantitative theoretical explanation to date. In previous theoretical attempts a much larger angle of about $50^{\circ}$ is usually found \cite{poulin98,smalyukh05}. This number is based upon calculations where one employs the electrostatic analog of the Boojum defect topology \cite{poulin98}. In fact, as stated explicitly by Poulin and Weitz ``This theoretical value is different from the experimentally observed value for $\theta$ $\ldots$ since the theory is a long-range description that does not account for short-range effects'' \cite{poulin98}. Another motivation for our work is the more recent experimental observation that between a pair of colloids in a nematic host repulsive and attractive forces act depending on $\theta$ \cite{smalyukh05}. For example, at $\theta\approx30^{\circ}$  the colloids attract each other whereas at $\theta=0^{\circ}$ and $90^{\circ}$ repulsion between the colloids is observed.

To unravel the persisting discrepancy between theory and experiment we employ a combination of Monte Carlo (MC) simulations in the isothermal-isobaric ensemble, two-dimensional (2D) MC simulations in the canonical ensemble, classical density functional theory (DFT), concepts of finite-size scaling (FSS), and Landau-de Gennes (LdG) theory to investigate the effective interaction between a pair of spherical, chemically homogeneous colloids mediated by a nematic host phase.

To model the host phase we adopt the so-called Hess-Su model. In this model mesogen-mesogen interactions are described by an isotropic core where $\varepsilon$ and $\sigma$ set energy and length scale, respectively. Superimposed to the isotropic core are anisotropic attractions of respective strengths $\varepsilon_1\varepsilon$ and $\varepsilon_2\varepsilon$ where the dimensionless anisotropy parameters $2\varepsilon_{1}=-\varepsilon_{2}=0.08$ throughout this work. Under these conditions the Hess-Su model exhibits isotropic-nematic (IN) phase transitions \cite{giura14}.

The colloid-mesogen interaction is modeled via short-range repulsive interactions and an attractive Yukawa tail where we take its inverse Debye screening length $\lambda\sigma=0.50$ \cite{melle12}. Mesogens at the colloids' surfaces are anchored in a locally planar fashion. This setup is then placed between structureless, planar solid substrates separated by a distance $s_{\mathrm{z}}=24\sigma$. Mesogens at the substrates are anchored such that their longer axes point along the $x$-axis $\hat{e}_{\mathrm{x}}$. Under these conditions a Boojum defect topology emerges at a single, isolated colloid. Colloids are immersed in the host phase such that their center-to-center distance vector is given by $\vec{r}_{12}=\left(x_{12},y_{12},0\right)$.

We employ dimensionless units, that is length is given in units of $\sigma$, energy in units of $\varepsilon$, and temperature in units of $\varepsilon/k_{\mathrm{B}}$ ($k_{\mathrm{B}}$ Boltzmann's constant). Other derived units are then expressed as combinations of these basic ones as usual \cite{melle12}. In particular, we set temperature $T=0.90$ and pressure $P=1.80$ such that the host phase is nematic at a mean number density $\rho\approx0.90$. The hard-core radius of each colloid is $r_{0}=3.00$. Other conditions of the MC simulations are exactly the same as in Ref.~\citealp{melle12} where additonal details of the model can also be found.

\begin{figure}[htb]
\centering
\epsfig{file=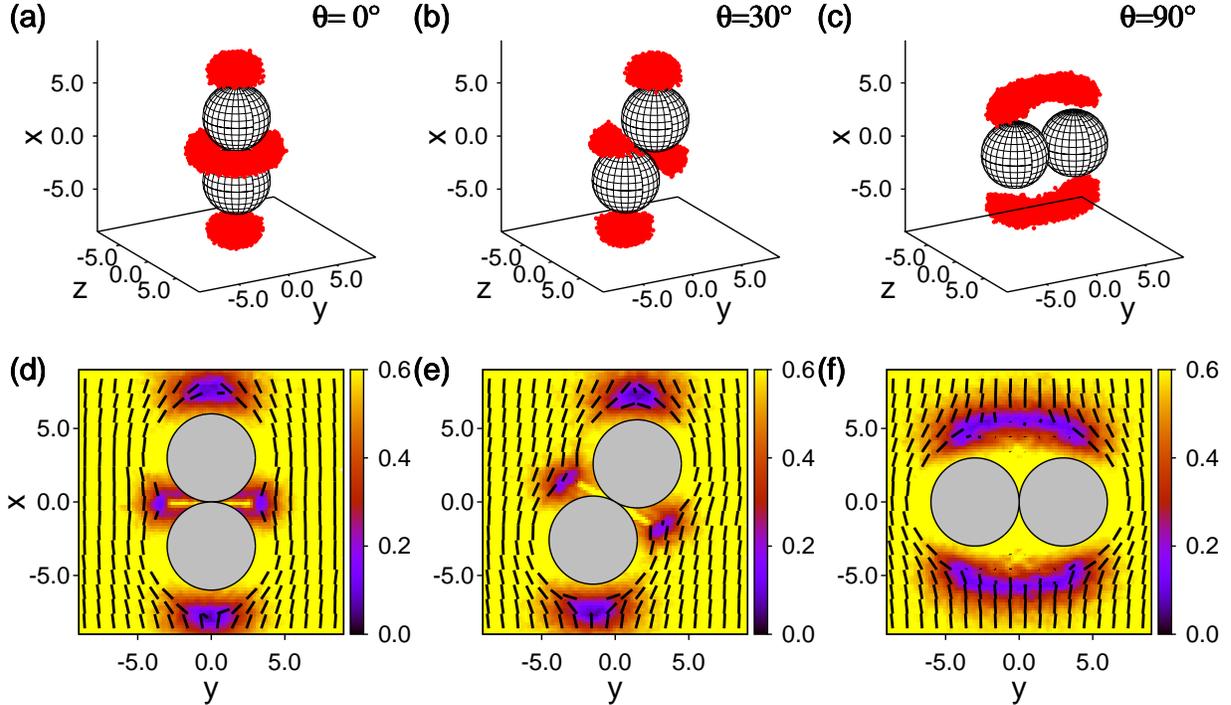,width=\linewidth}
\caption{(Color online) (a)--(c) Defect topologies for a colloidal pair with locally planar surface anchoring of mesogens separated by $\vec{r}_{12}$; $\cos\theta=\vec{r}_{12}\cdot\hat{n}_{0}/r_{12}$, $\hat{n}_{0}\cdot\hat{e}_{x}=1$, and $r_{12}=\left\vert\vec{r}_{12}\right\vert$. (d)--(f) As (a)--(c) but projected onto the $x$--$y$ plane. Attached color bars give $S\left(x,y\right)$ and dashes indicate $\hat{n}\left(\vec{r}\right)$.}\label{fig:config}
\end{figure}

Results of our MC simulations shown in Fig.~\ref{fig:config}(a) indicate that parts of the Boojum defects interact forming a torus. As $\theta$ increases, the torus is ``ripped apart'' [Fig.~\ref{fig:config}(b)]. Eventually, a handle-like defect topology emerges at $\theta\simeq90^{\circ}$ [Fig.~\ref{fig:config}(c)]. 

Defect regions around the colloids are visualized by shading them if the local nematic order parameter $S\left(\vec{r}\right)\le0.20$. We obtain $S\left(\vec{r}\right)$ numerically as the largest eigenvalue of the local alignment tensor  \cite{melle12}. The eigenvector $\hat{n}\left(\vec{r}\right)$ associated with $S\left(\vec{r}\right)$ is the director field.

The latter is illustrated in Figs.~\ref{fig:config}(d)--\ref{fig:config}(f). The plots indicate relatively localized regions of low $S\left(\vec{r}\right)$ in the vicinity of the colloids and that $\hat{n}\left(\vec{r}\right)$ is bent in ways that depend on the specific defect topology (i.e., on $\theta$).

Naturally, the reduction of $S\left(\vec{r}\right)$ and the bending of $\hat{n}\left(\vec{r}\right)$ causes the free-energy density $f\left(\vec{r}\right)$ of the system to increase locally relative to that of the pure host phase without the colloids. Consequently, we adopt
\begin{eqnarray}\label{eq:df}
\Delta f\left(\vec{r}\right)\!&=&\!
A\left(T,\rho\right)S^{2}\left(\vec{r}\right)\!+\!
B\left(T,\rho\right)S^{3}\left(\vec{r}\right)\!+\!
C\left(T,\rho\right)S^{4}\left(\vec{r}\right)\nonumber\\
&&+\frac{K}{2}
\left\lbrace
\left[\vec{\nabla}\cdot\hat{n}\left(\vec{r}\right)\right]^{2}+
\left[\vec{\nabla}\times\hat{n}\left(\vec{r}\right)\right]^{2}
\right\rbrace-f_{0}
\end{eqnarray}
where the first three terms on the right side correspond to a local LdG free-energy density $f_{\mathrm{LdG}}\left(\vec{r}\right)$, $f_{0}=AS^{2}+BS^{3}+CS^{4}$ is the LdG free-energy density obtained under the same thermodynamic conditions but in the absence of the colloids, and $\Delta f_{\mathrm{LdG}}\left(\vec{r}\right)\equiv f_{\mathrm{LdG}}\left(\vec{r}\right)-f_{0}$. Coefficients $A$, $B$, and $C$ are coefficients in the LdG expansion and $S$ is the global nematic order parameter.

The two terms on the second line of Eq.~(\ref{eq:df}) correspond to the local Frank free-energy density $f_{\mathrm{el}}\left(\vec{r}\right)$ that accounts for elastic distortions of the director field where $K$ is an elastic constant. We consider here the so-called one-constant approximation in which it is assumed that splay, twist, and bend deformations of $\hat{n}\left(\vec{r}\right)$ contribute equally to $f_{\mathrm{el}}\left(\vec{r}\right)$. It has recently been shown \cite{stieger14} that the one-constant approximation is an excellent approximation for the present model system because of the small aspect ratio of the mesogens. Under the present thermodynamic conditions, $K=1.66$. We then obtain $f_{\mathrm{el}}$ by numerically differentiating $\hat{n}\left(\vec{r}\right)$ \cite{melle12}.

We assume that both the local LdG contribution in Eq.~(\ref{eq:df}) and $f_{0}$ are governed by the same set $A$, $B$, and $C$. Moreover, $\Delta f\left(\vec{r}\right)$ in Eq.~(\ref{eq:df}) does not account for either fluctuations in $S\left(\vec{r}\right)$ or $\hat{n}\left(\vec{r}\right)$ and therefore constitutes a mean-field expression. Notice also that using in Eq.~(\ref{eq:df}) $S\left(\vec{r}\right)$ and $\hat{n}\left(\vec{r}\right)$ from MC is advantageous because then both quantities correspond to an equilibrium situation. Conventionally, $S\left(\vec{r}\right)$ and $\hat{n}\left(\vec{r}\right)$ are treated as variational functions in the {\em ansatz} in Eq.~(\ref{eq:df}) which bears the risk that the numerical minimization of the functional $\Delta f\left[S\left(\vec{r}\right),\hat{n}\left(\vec{r}\right)\right]$ may miss the true equilibrium solution.

To use Eq.~(\ref{eq:df}), $A$, $B$, and $C$ are required. Whereas these quantities are notoriously difficult to compute for reasons described by Eppenga and Frenkel a long time ago \cite{eppenga84}, Gupta and Ilg have devised a new approach that works reliably for mesogens with a relatively large aspect ratio \cite{gupta13}. In practice, however, we observed that the method of Gupta and Ilg does not work well for our model fluid where mesogens have a rather small aspect ratio of only $1.26$.

Because Eq.~(\ref{eq:df}) constitutes a mean-field expression we resort to mean-field DFT alternatively where \cite{giura13}
\begin{equation}\label{eq:deltafor}
\beta\Delta f_{\mathrm{or}}=
\rho\int\limits_{-1}^{1}\rmd x\,
\overline{\alpha}\left(x\right)
\ln\left[\overline{\alpha}\left(x\right)\right]+
\rho^{2}\sum\limits_{\substack{l=2\\l\text{ even}}}^{\infty}S_{l}^{2}u_{l}
\end{equation} 
is the difference in free-energy density of the nematic relative to the isotropic phase. In Eq.~(\ref{eq:deltafor}), $x=\cos\vartheta$ where $\vartheta$ is the azimuthal angle, $\beta=1/k_{\mathrm{B}}T$, $\overline{\alpha}\left(x\right)$ is the orientation distribution function, and members of the set $\left\lbrace u_{l}\right\rbrace$ account for the contribution of anisotropic mesogen-mesogen interactions to the free-energy density. Because of the uniaxial symmetry of the nematic phase we expand 
\begin{equation}\label{eq:xpnd}
\overline{\alpha}\left(x\right)=
\frac{1}{2}+
\sum\limits_{\substack{l=2\\l\text{ even}}}^{\infty}
\frac{2l+1}{2}S_{l}P_{l}\left(x\right)\equiv
\frac{1}{2}+\xi\left(x\right)
\end{equation}
in terms of Legendre polynomials $\left\lbrace P_{l}\left(x\right)\right\rbrace$.   We assume $\hat{n}_{0}\cdot\hat{e}_{\mathrm{z}}=1$ and $0\le S_{l}\le1$ are order parameters. 

We then insert the expression on the far right side of Eq.~(\ref{eq:xpnd}) into Eq.~(\ref{eq:deltafor}) and expand the integrand in terms of $\xi$ around $\xi=0$ (i.e., at the IN phase transition). Retaining in this expansion only the leading term of $\xi$ for $l=2$ and neglecting terms proportional to $S_{2}^{n}$ ($n\ge5$) allows us to rewrite Eq.~(\ref{eq:deltafor}) as
\begin{equation}\label{eq:dforfin}
\Delta f_{\mathrm{or}}=
a\left(\rho\right)
\left(T-
T^{\ast}
\right)S_{2}^{2}-
\frac{8\rho k_{\mathrm{B}}T}{105}S_{2}^{3}+\frac{4\rho k_{\mathrm{B}}T}{35}S_{2}^{4}
\end{equation}
where $a\left(\rho\right)=2\rho k_{\mathrm{B}}/5$ and $T^{\ast}=-5\rho u_{2}/2k_{\mathrm{B}}$ is the temperature at which the nematic phase becomes thermodynamically stable. Assuming that $S=S_2$ we equate terms of equal power in $S$ in $f_{0}$ and Eq.~(\ref{eq:dforfin}) which yields molecular expressions for the LdG constants $A$, $B$, and $C$. In particular, $A$ changes sign at $T=T^{\ast}$, $B<0$, and $C>0$ as they must at a first-order phase transition \cite{degennes95}.

One also notices that the value of $T^{\ast}$ depends on $u_{2}$ where the precise form of $u_{2}$ is a consequence of the level of sophistication at which pair correlations are treated within mean-field DFT \cite{giura14}. For example, at simple mean-field (SMF) level, $u_{2}=-32\pi\varepsilon_{1}\varepsilon\sigma^{3}/15$ is a constant. At the more elaborate modified mean-field (MMF) level, $u_{2}$ becomes a function of $T$ [see Eqs.~(3.7) and (3.8) of Ref.~\citealp{giura14}]. It turns out that at SMF level, $T^{\ast}$ is underestimated whereas at MMF level it is overestimated.

To overcome this problem we determine $T^{\ast}$ via FSS. Following Ref.~\citealp{greschek11} we first calculate the coexistence temperature $T_{\mathrm{IN}}\simeq1.02$ at the IN phase transition. It is given as the intersection of the second-order Binder cumulants of $S$ for different system sizes \cite{greschek11}. From the expression  $T^{\ast}=T_{\mathrm{IN}}-2B^{2}/9aC$ \cite{senbetu82} and using $B$, $C$, and $a$ from DFT, $T^{\ast}$ can easily be determined. Notice also that $S_{\mathrm{IN}}=-2B/3C=\frac{4}{9}$ irrespective of $T_{\mathrm{IN}}$ \cite{degennes95} whereas MMF DFT predicts this value of $S_{\mathrm{IN}}$ only to be a threshold reached for sufficiently high $T_{\mathrm{IN}}$ (Fig.~2 of Ref.~\citealp{giura14}) thus pointing to a certain deficiency of LdG theory.

\begin{figure}[htb]
\centering
\epsfig{file=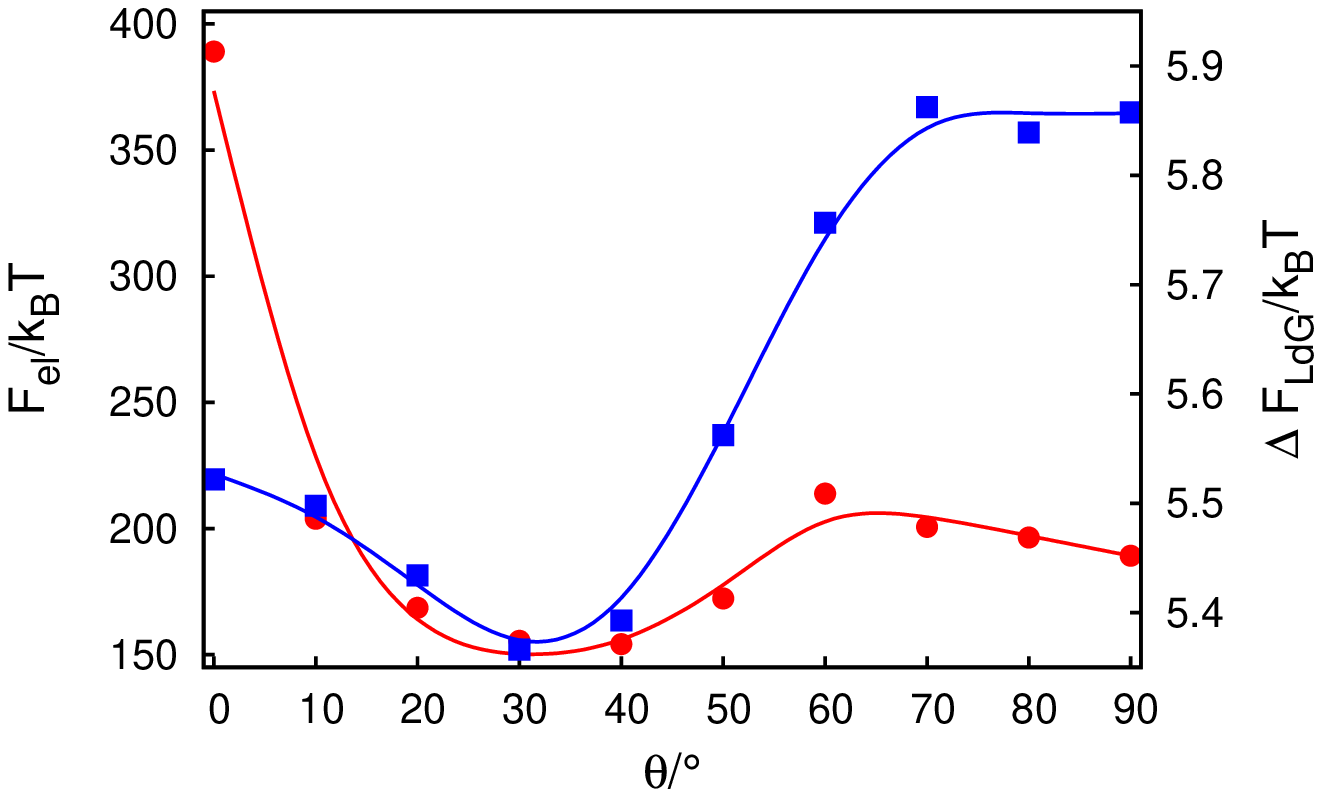,width=\linewidth}
\caption{(Color online) $\mathcal{F}_{\mathrm{el}}=\int\rmd\vec{r}\,f_{\mathrm{el}}\left(\vec{r}\right)$ (\red{\fullcircle}) (left ordinate) and $\Delta\mathcal{F}_{\mathrm{LdG}}=\int\rmd\vec{r}\,\Delta f_{\mathrm{LdG}}\left(\vec{r}\right)$ (\blue{\fullsquare}) (right ordinate) as functions of $\theta$ for $r_{12}=2r_{0}$ (see Fig.~\ref{fig:config}).}\label{fig:energy}
\end{figure} 

Plots of $\mathcal{F}_{\mathrm{el}}$ and $\Delta\mathcal{F}_{\mathrm{LdG}}$ in Fig.~\ref{fig:energy} illustrate the impact of a colloidal pair on the free energy of the host phase. Both quantities vary nonmonotonically with the angle $\theta$ and exhibit minima at $\theta\simeq30^{\circ}$ in agreement with the experimental findings of Poulin and Weitz \cite{poulin98}. Because deformations of $\hat{n}\left(\vec{r}\right)$ cost free energy, $\mathcal{F}_{\mathrm{el}}>0$.  Similarly, the presence of the colloids reduces $S\left(\vec{r}\right)$ such that in some regions $S>S\left(\vec{r}\right)$ (see Fig.~\ref{fig:config}). Because the host phase without the colloids is deep in the nematic phase, $f_{0}<0$ such that $\Delta\mathcal{F}_{\mathrm{LdG}}>0$ as well.

That both $\mathcal{F}_{\mathrm{el}}$ and $\Delta\mathcal{F}_{\mathrm{LdG}}$ become minimal at about the same $\theta$ indicates that destortions of $\hat{n}\left(\vec{r}\right)$ and a local reduction of nematic order are coupled. However, deformations of $\hat{n}\left(\vec{r}\right)$ turn out to be more important than reduction of nematic order because $\mathcal{F}_{\mathrm{el}}$ exceeds $\Delta\mathcal{F}_{\mathrm{LdG}}$ by between one and two orders of magnitude over the entire range of $\theta$'s. This conclusion is drawn on the basis of plots in Fig.~\ref{fig:energy} and by noticing that for both curves the ground state is the same, namely $\hat{n}\left(\vec{r}\right)=\hat{n}_{0}$ ($\mathcal{F}_{\mathrm{el}}=0$) and $S\left(\vec{r}\right)=S$ ($\Delta\mathcal{F}_{\mathrm{LdG}}=0$).

\begin{figure}[htb]
\centering
\epsfig{file=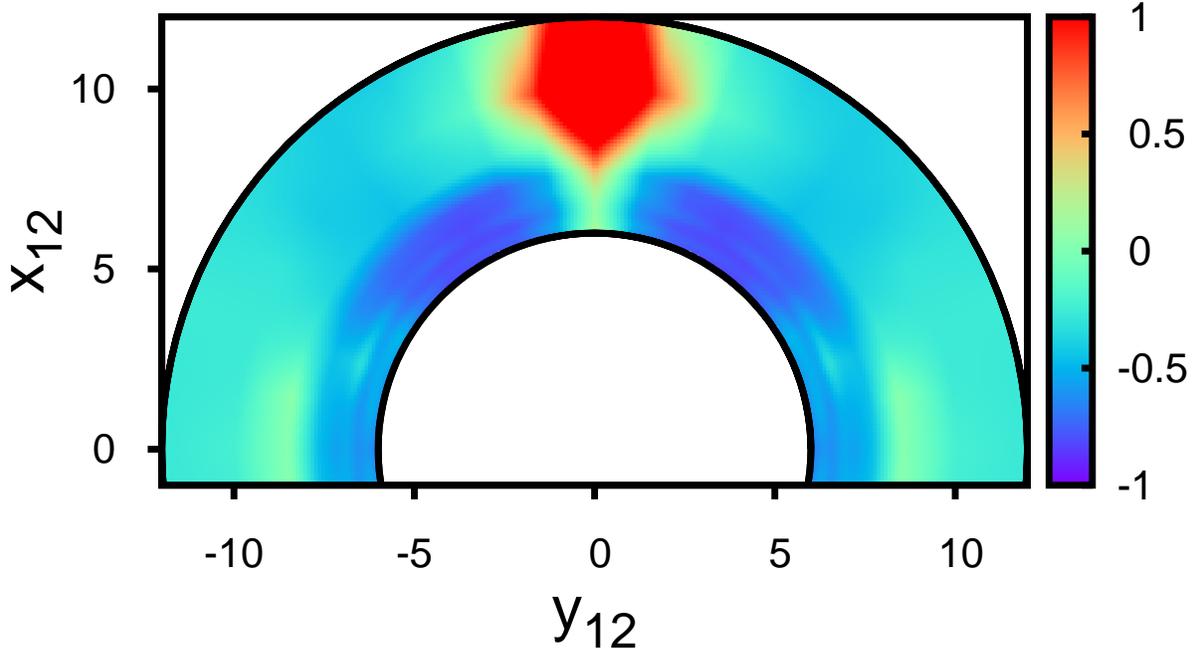,width=\linewidth}
\caption{(Color online) $\Delta\mathcal{F}_{\mathrm{eff}}/\Delta\mathcal{F}_{\mathrm{B}}$ as a function of relative positions of the colloids in the $x$--$y$ plane (see attached color bar). The white semicircle at the center represents a reference colloid.}\label{fig:auquiplot}
\end{figure}

Results presented in Fig.~\ref{fig:energy} have been obtained for two colloids in contact with each other. However, the general physical picture reflected by Fig.~\ref{fig:energy} is preserved if besides $\theta$, $r_{12}$ is varied, too. To that end we realize from Eq.~(\ref{eq:df}) that $\lim_{r_{12}\to\infty}\Delta f\left(\vec{r}\right)=2\Delta f_{\mathrm{B}}\left(\vec{r}\right)$ where $\Delta f_{\mathrm{B}}\left(\vec{r}\right)$ is the local free energy density of two isolated Boojum defects relative to the same ground state used above. Taking $\Delta\mathcal{F}_{\mathrm{B}}=\int\rmd\vec{r}\,\Delta f_{\mathrm{B}}\left(\vec{r}\right)$ allows us to introduce $\Delta\mathcal{F}_{\mathrm{eff}}\equiv\mathcal{F}_{\mathrm{el}}+\Delta\mathcal{F}_{\mathrm{LdG}}-2\Delta\mathcal{F}_{\mathrm{B}}$ as the {\em effective} potential acting between a pair of colloids and mediated by the nematic host.

A map of $\Delta\mathcal{F}_{\mathrm{eff}}$ in Fig.~\ref{fig:auquiplot} shows that for $\theta=0^{\circ}$, $\Delta\mathcal{F}_{\mathrm{eff}}$ is strongly repulsive in a relatively localized region. This is a consequence of the merger of parts of the Boojum defect illustrated by Figs.~\ref{fig:config}(a) and \ref{fig:config}(d). In agreement with plots in Fig.~\ref{fig:energy} we see that 
$\Delta\mathcal{F}_{\mathrm{eff}}$ is attractive if $r_{12}$ is sufficiently small where the absolute minimum of $\Delta\mathcal{F}_{\mathrm{eff}}$ is found at $\theta\approx30^{\circ}$.

One also notices from Fig.~\ref{fig:auquiplot} a small repulsive barrier in $\Delta\mathcal{F}_{\mathrm{eff}}$ as $\theta$ approaches $90^{\circ}$ and $7\lesssim r_{12}\lesssim10$. Hence, a pair of colloids at $\theta\approx0^{\circ}$ and at sufficiently large $r_{12}$ and $75^{\circ}\lesssim\theta\lesssim90^{\circ}$ would repel each other whereas those forming an angle of $\theta\approx30^{\circ}$ would attract each other. These findings are in excellent agreement with experimental observations [Fig.~2(b) of Ref.~\citealp{smalyukh05}].

\begin{figure}[htb]
\centering
\epsfig{file=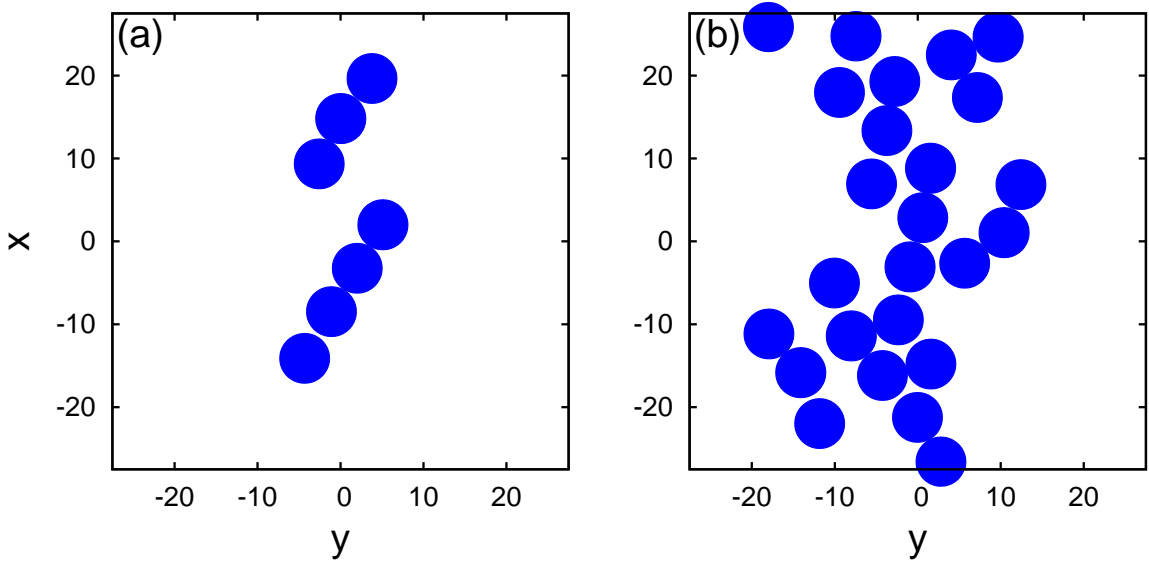,width=\linewidth}
\caption{(Color online) 2D MC configurations ($\hat{n}_{0}\cdot\hat{e}_{\mathrm{x}}=1$). (a) $\phi=N_{\mathrm{coll}}\pi r_{0}^{2}/s_{\mathrm{x}}s_{\mathrm{y}}=0.065$, (b) $\phi=0.234$ ($s_{\mathrm{x}}=s_{\mathrm{y}}=50$).}\label{fig:self_assembly}
\end{figure}

Taking $\Delta\mathcal{F}_{\mathrm{eff}}$ as an effective, pairwise additive potential we perform standard Metropolis 2D MC simulations of $N_{\mathrm{coll}}$ colloids modeling the nematic host phase implicitly. Technically, $\Delta\mathcal{F}_{\mathrm{eff}}$ is stored at nodes of a regularly spaced grid in the $x$--$y$ plane; the actual value of $\Delta\mathcal{F}_{\mathrm{eff}}$ at $\vec{r}_{12}$ is obtained by bilinear interpolation between the four nearest nodes. The simulations are carried out in the canonical ensemble. Results in Fig.~\ref{fig:self_assembly}(a) show that at low packing fraction $\phi$ the colloids tend to form linear chains of an angle of about $30^{\circ}$ with $\hat{n}_{0}$. At higher $\phi$ the snapshot in Fig.~\ref{fig:self_assembly}(b) reveals more extended two-dimensional structures. Plots in both parts of Fig.~\ref{fig:self_assembly} are in excellent qualitative agreement with experimental findings (see Fig.~1 of Ref.~\citealp{smalyukh05}).

To summarize we used a combination of MC simulations, FSS, and mean-field DFT to compute the effective interaction potential between a pair of colloids immersed in a nematic liquid crystal. The colloids are chemically homogeneous and anchor mesogens in a locally planar fashion at their surface. On accound of the mismatch between this local alignment and $\hat{n}_{0}$ a Boojum defect topology emerges at an isolated colloid. If two such colloids approach each other the Boojum defects interact such that the precise topology changes with the angle $\theta$ formed between the distance vector connecting the centers of the colloidal pair and $\hat{n}_{0}$.

As a result of the topological change repulsive and attractive effective interactions arise. These are dominated by the distortion of $\hat{n}\left(\vec{r}\right)$ whereas the accompanying reduction of local nematic order is negligible. Most notably, the distribution of regions in which the effective interaction potential $\Delta\mathcal{F}_{\mathrm{eff}}$ is attractive or repulsive matches experimental results reported by Smalyukh {\em et al.} despite their much larger colloids \cite{smalyukh05}. 

It is particularly gratifying that the most favorable angle we find is $\theta\approx30^{\circ}$ in agreement with the work by Poulin and Weitz \cite{poulin98} and Smalyukh {\em et al.} \cite{smalyukh05}. Our work therefore offers the first quantitative theoretical explanation of earlier experimental observations. Moreover, we show that it is the relatively short-range effects that are responsible for the observed attraction and repulsion between nematic colloids thereby confirming the earlier conjecture by Poulin and Weitz \cite{poulin98}.  

\acknowledgments
We acknowledge financial support from {\em Deutsche Forschungsgemeinschaft} through the International Graduate Research Training Group 1524. S.~S. and M.~M. are grateful for discussions with Prof. C.~K.~Hall (NCSU).

\end{document}